# Leveraging power of deep learning for fast and efficient elite pixel selection in time series SAR interferometry

Ashutosh Tiwari[1]*, Nitheshnirmal Sadhashivam[2], Leonard O. Ohenhen[2], Jonathan Lucy[2] and Manoochehr Shirzaei[2]

Texas A&M Agrilife Research Center, Corpus Christi, TX, USA
Virginia Tech National Security Institute, Blacksburg, VA, USA
Corresponding author: Ashutosh Tiwari (ashutoshtiwari796@gmail.com)

**Abstract-** **This study proposes a new convolutional long short-term memory (ConvLSTM) based architecture for selection of elite pixels (i.e., less noisy) in time series interferometric synthetic aperture radar (TS-InSAR). The model utilizes the spatial and temporal relation among neighboring pixels to identify both persistent and distributed scatterers. We trained the model on ~20,000 training images (interferograms), each of size 100 × 100 pixels, extracted from InSAR time series interferograms containing both artificial features (buildings and infrastructure) and objects of natural environment (vegetation, forests, barren or agricultural land, water bodies). Based on such categorization, we developed two models, tailormade to detect elite pixels in urban and coastal sites. Training labels were generated from elite pixel selection outputs generated from the wavelet-based InSAR (WabInSAR) software. We used 4 urban and 7 coastal sites for training and validation respectively, and the predicted elite pixel selection maps reveal that the proposed models efficiently learn from WabInSAR-generated labels, reaching a test accuracy of 94%. The models accurately discard pixels affected by geometric and temporal decorrelation while selecting pixels corresponding to urban objects and those with stable phase history unaffected by temporal and geometric decorrelation. The density of pixels in urban areas is comparable to and higher for coastal areas than WabInSAR outputs. With significantly reduced time computation (order of minutes) and improved density of elite pixels, the proposed models can efficiently process long InSAR time series stacks and generate deformation maps quickly, making the time series InSAR technique more suitable for varied (non-urban and urban) terrains and unaddressed land deformation applications.**

## I. INTRODUCTION

Time series interferometric synthetic aperture radar (TS-InSAR) has been used extensively over the last decade to monitor natural and anthropogenic surface deformation owing to its high density of observations with reliable phase change information [1], [2], [3][4], [5], [6], [7]. With a further rise in the datasets thanks to the present (Sentinel-1, RADARSAT, ALOS PALSAR, TerraSAR-X) and upcoming (NISAR, Tandem-L, COSMO-SkyMed second generation, TerraSAR-3) SAR data missions [8], [9], [10], [11], [12], [13], [14], the processing requirements are and going to be more demanding. It is, hence, incumbent upon researchers, scientists, and engineers working on InSAR to devise efficient ways of data processing, transfer, and storage using advanced data structures, memory management, and information mining techniques. The overall TS-InSAR processing chain generally involves preprocessing the single look complex (SLC) SAR images (for orbit determination, multi-looking, burst removal for Sentinel-1) followed by co-registration, interferogram generation and partial topographic phase removal through an external digital elevation model [4], [7] [6], [15], [16], [17]. These steps yield a stack of differential interferograms ready for time series processing. Different methods are used to perform TS-InSAR, mostly chosen based on the phenomena investigated and the terrain characteristics [6], [7], [16], [18]. Subsequently, the methods perform the selection of elite (less-noisy) pixels, phase unwrapping, estimation and removal of nuisance phase, and displacement estimation to evaluate the spatiotemporal deformation pattern. The pixel selection step here is critical in governing the observation set for the succeeding steps. The density and quality of pixels vary due to different types of terrain, object characteristics (artificial or natural), geometrical and temporal decorrelation, and deformation behavior. Two types of scatterers are useful for interferometry, (i) persistent scatterers (PS) and (ii) distributed scatterers (DS), and we categorize these as elite pixels for this study too [16], [18].

With the requirement of spatio-temporal analysis of millions of pixels present in a differential interferogram stack, elite pixel selection consumes a significant chunk of processing time for most TS-InSAR processors [19], [20]. For methods working only with PS pixels, the quality and density of elite pixels also suffer in the case of non-urban areas, where PS pixels are rare [21], [22], [23]. The conventional pixel selection methods used by TS-InSAR processors typically require hours and sometimes days in case the computational resources are constrained. We see that artificial intelligence tools are proving invaluable for various fields of research, including earth observation studies, providing methods that are computationally efficient and accurate [24], [25], [26], [27], [28], [29], [30]. For InSAR studies too, we see research related to volcano detection [31], urban area deformation analysis [26], land subsidence [32], landslide zones and slope instability [33], [34], measurement pixel selection[35] [36], phase unwrapping [37], [38], deformation prediction [39] and post-processing deformation signals [40], [41], [42].

Motivated by the challenges present in elite pixel selection and the promising outcomes from AI-related methods, we propose a deep learning-based approach to perform elite pixel selection in the TS-InSAR processing chain for this study. Though several existing semantic segmentation models developed in the field of computer vision [25], [43], [44] are promising, we observe that InSAR phase change images (interferograms) belonging to diverse natural objects with geometrical and topographical variations pose different challenges while reusing or retraining these models. Studies showing pixel-wise labeling of remote sensing images [45], [46] show promising results, but SAR interferograms with pixels affected by diverse scattering mechanisms and response characteristics require dedicated model development and hyperparameter learning. This study aims to develop deep learning architectures automating the selection of elite (both PS and DS) pixels in TS-InSAR with reduced computational requirements and comparable or better selection quality. Deep learning models based on convolutional neural network and convolutional LSTM were proposed in [1], providing a time efficient approach to detect PS pixels. But the models were trained on StaMPS based PS pixel selection labels, which are affected by low density in non-urban areas. Further, the models could only work for a specific number of time steps. We hereby propose building a model named *ConvLSTM for InSAR pixel selection* (CIPS), using convolutional long short term memory network allowing the selection of both PS and DS pixels, and allow any number of interferograms to be used for pixel selection. Further, we develop multiple deep learning models, each suited to a particular type of terrain rather than one generalized model satisfying all study sites. It may be debatable that one generalized model can be built and replaced with multiple specific models for pixel selection. We however believe that for the problem statement addressed here, the cost of misinterpretation of deformation pattern and the reliability of reported deformation estimates is more significant than having an ease of using a single generalized model.

The paper is organized as follows. Section 2 details the dataset preparation for training the deep learning models. Section 3 describes the proposed CIPS model architecture. Section 4 presents the results obtained and discussions, followed by conclusions in section 5.

## II. STUDY SITES AND DATASETS

Based on the diversity of the acquisition geometry, terrain characteristics, deformation behavior, and user interests, we focus this study on two different types of study sites, urban and coastal. To cover the whole spectrum of scatterers, we chose some of the coastal sites encompassing non-urban features like hills, barren areas, forests, and water bodies. We seek to build separate deep learning models for urban and coastal sites characterized by different geometric effects (foreshortening, layover and shadow), land use/land cover characteristics (forest, urban, agricultural, barren land, etc.) affected by varied deformation types (linear or non-linear) and phenomena (land subsidence, uplift, landslides, earthquakes, volcanic activity, etc.). Fig. 1(C) shows a schematic view of the object backscatter responses received for different types of terrain. Fig. 1(D) shows that urban areas and man-made infrastructure receive responses like that of PS, while agricultural fields give a DS-like response. Coastal sites also have a human population with a few man-made structures and natural objects, yielding both PS- and DS-like responses. Forest areas generally contain poor scatterers, and hilly areas retain PS in case of foreshortening, which results in noisy scatterers for layover and shadowing affected areas. We use four urban and six coastal sites for training and two sites each for testing the urban and coastal models. Fig. 1(A) and 1(B) show the urban and coastal sites, respectively. The first model, named the 'urban model,' uses four cities in the United States for training, spread well apart having different distribution of buildings and infrastructure, and having different deformation behaviors altogether, as evidenced by [47], [48], [49], [50], [51], [52]. We chose the United States East Coast to train the second model and named it the 'coastal model' for simplicity here. The East coast, with ~3500 km coastline, has varying terrain characteristics containing forests and agricultural lands, and a few settlements inland, and is affected by land subsidence [53], [54], [55], [56], [57]. More information on the dataset parameters is provided in Supplementary Tables I A and I B.

We divide each differential interferogram into patches of 100 by 100 pixels, making them suitable for memory management during the training of the deep learning models and also increasing the sample size. Effectively, the models are trained on ~20,000 interferometric images (each of size 100 by 100), ensuring enough training samples. We use labels of elite (persistent and distributed) pixels extracted from the Wavelet-based InSAR (WabInSAR) processing chain proposed in [7], and improved by [17]. WabInSAR method uses a statistical approach to perform elite pixel selection, initially selecting PS candidates based on a specified threshold on amplitude dispersion index $D_A$, as shown in equation 1.1, where $\sigma_a$ and $\mu_a$ denote the standard deviation and mean values of pixel amplitude for the time series.

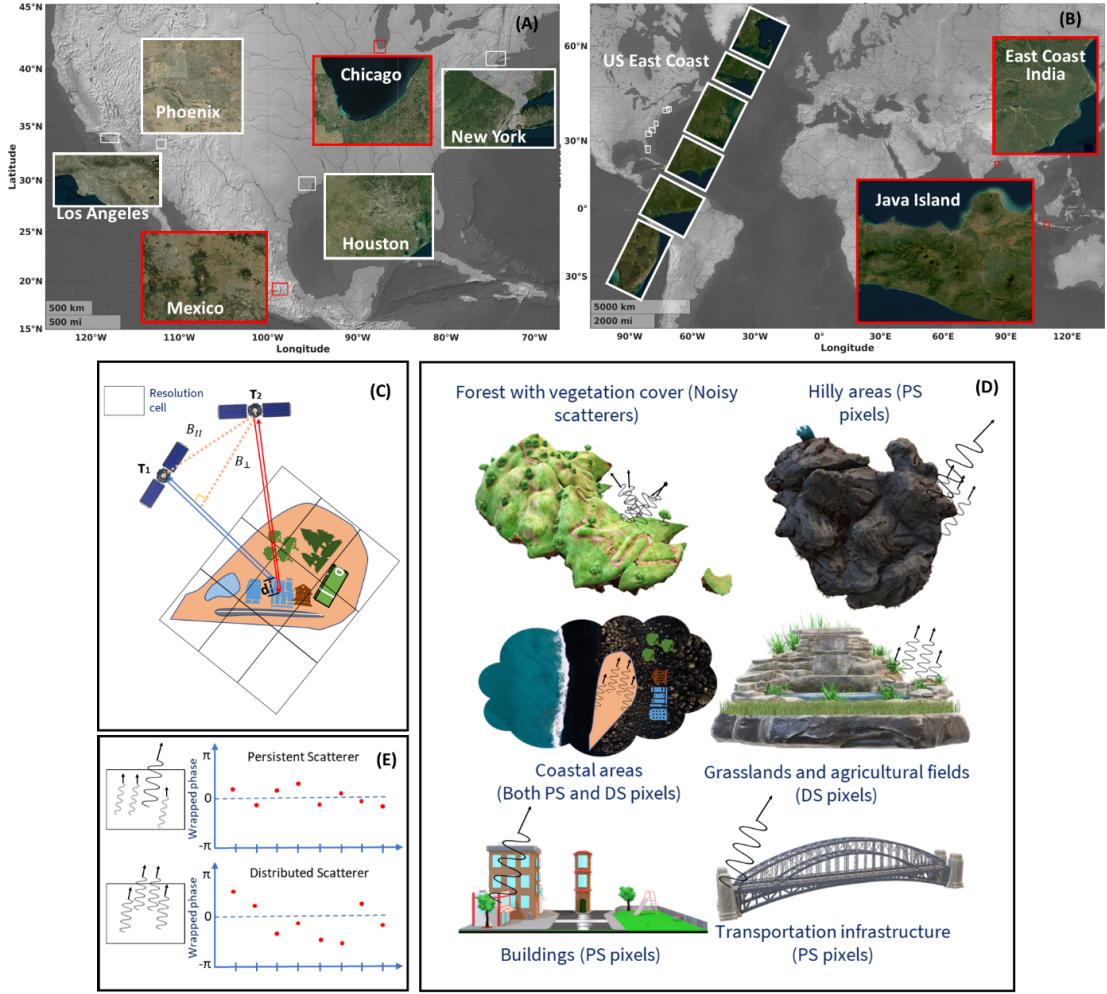

**Fig. 1**: A schematic view of target responses in SAR interferometry. (A) and (B) show study sites used for training the urban and coastal sites, respectively. White and red border rectangles denote training and test datasets, respectively. (C) and (D) show InSAR geometry and the scattering response mechanisms for different terrains. (E) shows a sample temporal response from persistent and distributed scatterers.

Later, DS candidates are selected using the coherence dispersion threshold $D_c$ as shown in equation 1.2, where $\sigma_c$ and $\mu_c$ represent the standard deviation and mean of the pixel coherence values for the time series. The DS candidates go under statistical evaluation with the PS pixels. The method subsequently forms voronoi cells containing PS candidates and identify their neighboring DS candidates, the cell size is decided based on the density of PS pixels. The DS candidates that satisfy statistical similarity with a PS candidate in a cell based on the Fisher's test of distribution similarity (equation 2) are considered elite pixels and are retained along with PS candidates to act as the final set of elite pixels. In equation 2, $F$ stands for the $F$-statistic value from the Fisher's distribution, $\sigma_{PS}$ stands for the amplitude standard deviation of the PS pixel over time, and $\sigma_{DS}^j$ stands for the standard deviation of the $j^{th}$ DS candidate pixel within the Voronoi diagram. If the computed value of $F$ turns out to be greater than the theoretical value, the pixel is included as an elite pixel and rejected otherwise. More details on this approach can be seen in [17].

$$D_A = \frac{\sigma_a}{\mu_a} \approx \sigma_{\emptyset_{noise}} \quad (1.1)$$

$$D_c = \frac{\sigma_c}{\mu_c} \quad (1.1)$$

$$F = \frac{(\sigma_{DS}^j)}{(\sigma_{PS})}, j=1..q \quad (2)$$

We divide the training set into a training and validation ratio of 70:30 for each training site. Most datasets we used in training contain about 80 to 150 time steps. However, for many research labs, it is challenging to process interferometric stacks with 100 time steps due to constrained computational resources compared to the humongous processing size requirements. To this end, we sample each interferometric stack to 25-30 time steps, with the model capable of accommodating stacks with different time steps. We also observe that the model works well with about 50 time steps, allowing different sampling rates for learning and prediction.

### III. PROPOSED DEEP LEARNING ARCHITECTURE

The proposed architecture is shown in Fig. 2. Prior to learning, we saw that the interferograms were divided into patches, each with a size of 100 by 100 pixels. Subsequently, the time series are sampled in fewer steps (25 to 30). The dimension of input passed to the architecture is $(n_s, n_t, w, h, f)$ where $n_s$ and $n_t$ represent the number of image samples and time steps respectively, $w$ and $h$ denote the image width and height respectively (100 each in this study), and $f$ denotes the number of features used for learning. The model consists of an input layer, two hidden convolutional LSTM layers named *convlstm*, a convolutional layer *conv* dedicated to spatial learning once the time series features are identified, and an output layer. To combat overfitting, the model encompasses a normalization layer between *convlstm₁* and *convlstm₂*, a batch normalization layer between *convlstm₂* and *conv*, and a dropout layer between *conv* and the last fully connected *FC* layers.

The structure of a *convlstm* cell is also shown in Fig. 2. Like an LSTM cell, it has three gates; forget, input and output, and a cell state. The operations of the cell state and the gates are similar, except for the fact that the convolution operation replaces each scalar multiplication present in an LSTM cell. This enables the *convlstm* layer spatially within a specified neighborhood (size of convolution kernel), and temporally with the basic LSTM structure. Equations 3.1 to 3.5 give the mathematical explanations of the three gates and the cell state. The response from the forget, input, and output gates obtained for time step *t* are represented as $fg_t$, $in_t$ and $out_t$ respectively. All these gates work with the output received from the previous time step $y_{t-1}$ and the input at the current time step $x_t$. A sigmoid activation is applied to the output of the three gates, with weights and biases different for each gate, denoted as $(w_{fg}, b_{fg})$, $(w_{in}, b_{in})$ and $(w_{out}, b_{out})$ for the forget, input, and output gates, respectively. Learning over iterations, the forget gate determines and discards unwanted information from the time series (equation 3.1).

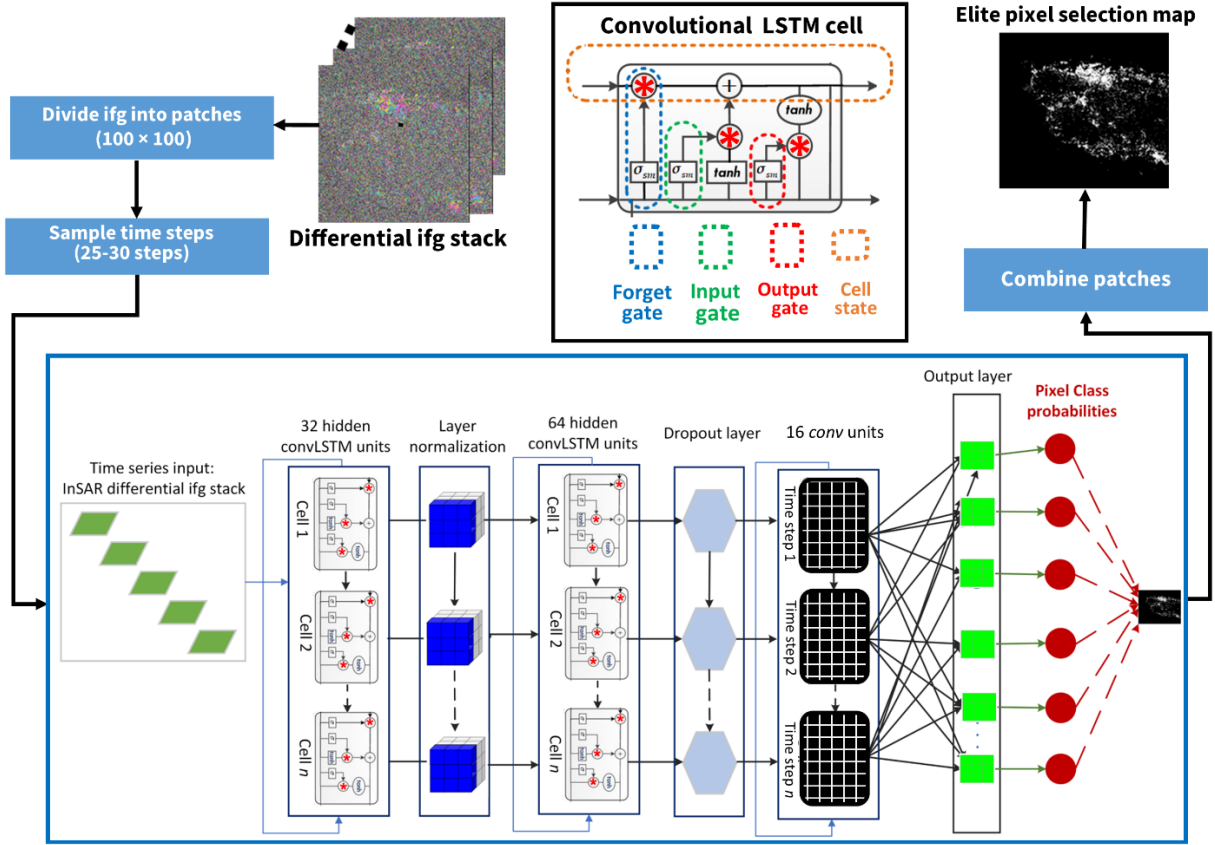

Fig. 2: Proposed method for elite pixel selection. Interferograms are divided into small patches of size 100 by 100 pixels and sampled in fewer time steps (25-30). This goes to the *CIPS* model for learning. Later, the output patches are combined to create the full elite pixel selection map for the input interferogram stack.

The input gate estimates and retains the useful information in the time series (equation 3.2). The cell uses the output of the forget and input gates to compute a predicted cell state. The final cell state is computed by adding the outputs received from convolving the present forget gate response $fg_t$ with the previous time step cell state $S_{t-1}$, and by convolving the present input gate response $in_t$ with $S_{t-1}$. This cell state acts as the memory of the *convlstm* cell and is passed over to the other cells in the layer to learn useful patterns in the time series. It uses a hyperbolic tangent activation *tanh* instead of sigmoid, and has its own weights and biases $(w_S, b_S)$ similar to the three gates. These get initialized randomly and are modified based on the loss values and learning over multiple epochs. The final output $y_t$ for time step $t$ is computed by convolving the output gate response with activated cell state response $tanh(S_t)$. This continues for the time series over a chosen number of epochs, and the weights and biases are tweaked to optimize the loss function. We use the $f_1$-*loss* as the loss function in this study for training to minimize the false positives and false negatives. Its computation is shown in Equation 4, where TP, FP and FN denote true positives, false positives and false negatives, respectively.

$$fg_t = \sigma_{sm}(w_{fg}[y_{t-1}, x_t] + b_{fg}) \quad \}Forget\ gate \quad (3.1)$$

$$\begin{aligned} in_t &= \sigma_{sm}(w_{in}[y_{t-1}, x_t] + b_{in}) \\ \tilde{S}_t &= tanh(w_S[y_{t-1}, x_t] + b_S) \end{aligned} \Big\} Input\ gate \quad (3.2)$$

$$S_t = fg_t * S_{t-1} + in_t * \tilde{S}_t \quad \}Cell\ state \tag{3.3}$$

$$\left.\begin{aligned}out_t &= \sigma_{sm}(w_{out}[y_{t-1}, x_t] + b_{out}) \\ y_t &= out_t * tanh(S_t)\end{aligned}\right\} Output\ gate \tag{3.4}$$

$$f_1 = 2 \times \left(\frac{precision \times recall}{precision + recall}\right)$$
$$precision = \frac{TP}{TP + FP} \quad recall = \frac{TP}{TP + FN} \tag{4}$$

The models in this study are trained using transfer learning. This implies that after running the model on one dataset, the model weights and biases are stored, and these act as initial weights and biases for the next dataset. The configuration of the proposed model is shown in Supplementary Table II, where *layernorm* and *batchnorm* denote the layer normalization and batch normalization layers (added to avoid overfitting), *relu* denotes the rectified linear unit activation, computed as $relu(x) = max(x, 0)$. The term *dropout* represents the Dropout layer, experimented with different ratios during transfer learning to achieve the best output. Hyperparameters, except for learning and decay rates, are kept the same for learning on different datasets. The hyperparameters used for the urban and coastal model learning are shown in Supplementary Table III. We trained each dataset with 70% points chosen randomly as training and 30 % as validation. The code and an example data for training the proposed model is available at https://github.com/ICCT-ML-in-geodesy/InSAR-based-pixel-selection.

## IV. RESULTS AND DISCUSSION

With multiple urban and coastal datasets trained with two different models, we predicted elite pixels for two urban and two coastal sites, noting the prediction time used per dataset. The prediction outputs for both models are shown in Fig. 3, where the left and right columns show pixels selected by the WabInSAR and CIPS model, respectively. Pixel densities for both urban and coastal datasets are comparable and higher for one coastal (Indian east coast) and one urban dataset (Chicago), as shown in Fig. 3(I) and Supplementary Table IV. The proposed model rejects temporally decorrelated objects like forest trees, water bodies, and snow while selecting man-made objects (buildings and other infrastructure) and agricultural land. The training time varies between 120 to 180 minutes. The prediction time is effectively the time required when the models are deployed to perform elite pixel selection, varying between 3 to 10 minutes as labeled over Fig. 3(B), (D), (F) and (H). We notice that the models outperform all the conventional InSAR processors in terms of computational efficiency, which consumes hours for pixel selection. Further, we achieve this efficiency while using full InSAR image frames and multiple combined frames for training and prediction.

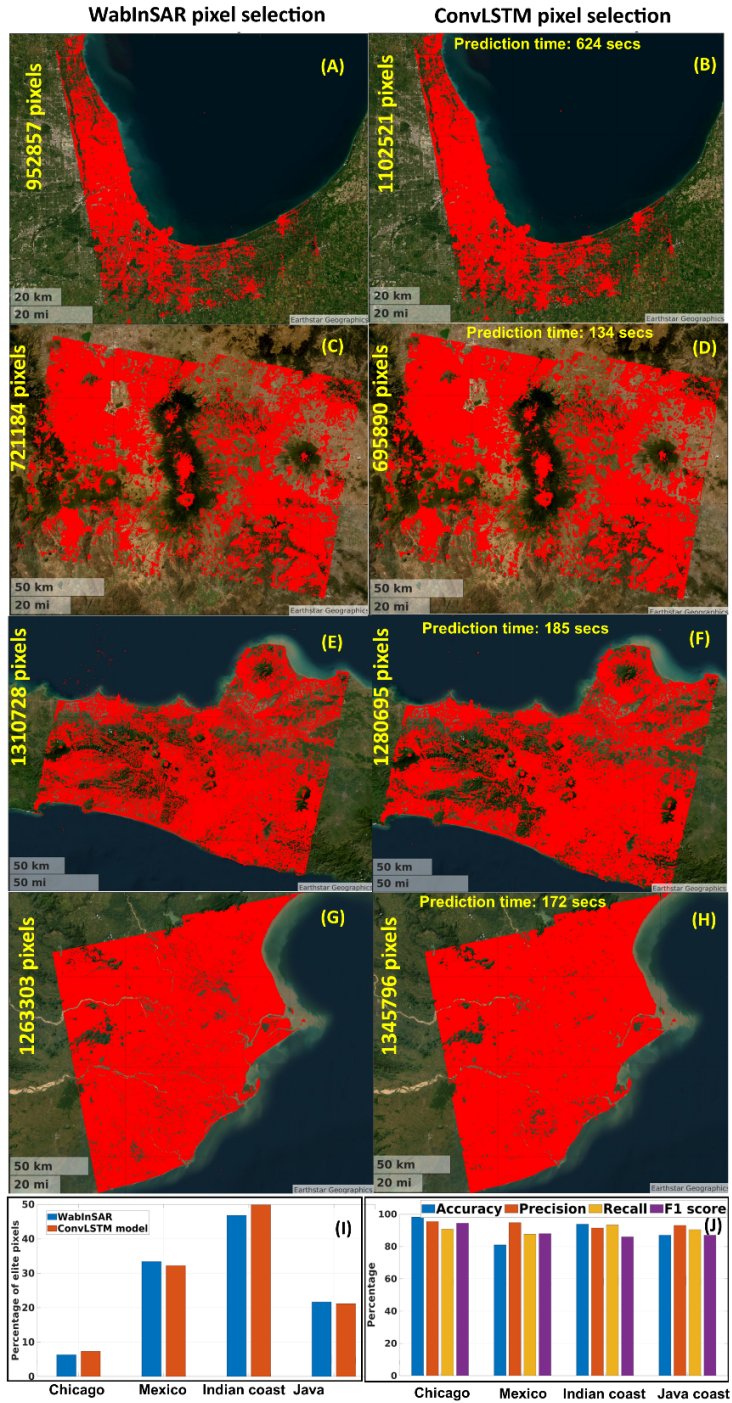

Fig. 3: Comparison of pixel selection results from conventional TS-InSAR software (WabInSAR) and the proposed ConvLSTM model. The Left and right columns show the results for the predicted urban and coastal sites with WabInSAR and ConvLSTM, respectively. Panels (A), (C), (E) and (G) show WabInSAR-based elite pixel selection for Chicago, Mexico, Java Island and Indian east coast, respectively. Panels (B), (D), (F) and (H) show corresponding CLSTM-ISS model-based selection for Chicago, Mexico, Java Island and Indian east coast, respectively. (I) shows a bar chart of the percentage of elite pixels selected for the test sites, and (J) shows a bar chart of performance metrics for the test sites.

We compute the quality of pixel selection using accuracy, precision, recall and $f_1$-score shown in equation 4 using the confusion matrices shown in Supplementary Table V. The metric outcomes are shown in Fig. 3 (J) and also tabulated in Supplementary Table VI. We observe that the prediction accuracy is always above 90 % and reaches 98 % for the Chicago dataset. We obtain good $f_1$-scores since we balanced false positives and false negatives while training. Embedding the proposed CIPS model into the WabInSAR processing chain, we executed the remaining steps, i.e. estimation and removal of nuisance phase components, phase unwrapping and displacement estimation. We perform this to understand the effect of the modified pixel selection over the final InSAR outcome. Fig. 4 shows a comparison of the one-dimensional line of sight velocities (1D-LOS) generated from the original WabInSAR processing (first column) and modified WabInSAR processing (second column), replacing the existing pixel selection step with the proposed method for the prediction datasets. As evident from the velocity maps, the velocity maps are similar, and the major deformation zones are seen in both outcomes. We further experimented with test points (random locations) in the test sites and compared the displacement time series. These test points are marked as white pentagons over Fig. 4 (B), (D), (F), and (H). Since the exact location of the test points did not have pixels selected by both methods, we computed the average displacement time series for pixels within 100 meters of the chosen location. Fig. 4 (I), (J), (K) and (L) show the compared time series for the test points. We see that the pattern is similar, while the magnitudes are slightly varying. Fig. S3 depicts the pixel differences in mean LOS velocities for the test sites. The differences are within 5 mm/year for panels (A) Chicago, (B) Mexico, and (C) Java Island, and within 10 mm/year for (D) Indian East coast, with most of the pixel differences below 2 mm/year. The pixels with higher differences however show similar deformation pattern, as evident from Fig. 4. We hence observe that the proposed model retains the quality of time series displacement. When choosing between urban and non-urban models, users should examine the geometry and terrain characteristics of the study area to decide which model is more suitable. The rule of thumb is that for a terrrain mainly comprising artificial features such as buildings, the urban model works better and vice versa. The models also account for outlying scattering mechanisms.

To evaluate the transferability of both the models (urban and coastal), we choose an urban site (Chicago) and a coastal site (Java Island) and predict the elite pixels using the coastal model on the urban site and vice versa. The results are added as Supplementary Table VII and Supplementary Fig. S2. We find that for the coastal study site, the density of the elite pixels is reduced to almost half when trained with the urban model (Supplementary Table VII). Further, we observe a rise in the density of elite pixels when we predict elite pixels in urban areas using the coastal model. In the latter case, the additional pixels are false positives and may affect the final deformation estimates. We see that the models efficiently select pixels over sites located in different parts of the globe, which are characterized by different terrain, geometry, and the presence of urban objects, showing their transferability. This further implies that the urban model can be used for performing pixel selection for study sites predominated with urban objects, while the coastal model can be used for sites characterized by non-urban objects. Compared with the model propoed in [35], this study broadens the applicabilities of the deep learning models. Existing models are mainly suitable for selecting permanent scatterer (PS) pixels, which are generally abundant in urban areas and scarce otherwise. In contrast, this study provides models that enable the selection of distributed scatterer (DS) pixels in addition to (PS), yielding an improved density of measurement points. This enhances the use of the proposed models for monitoring deformation in diverse settings, increasing the use of deep learning in SAR interferometry.

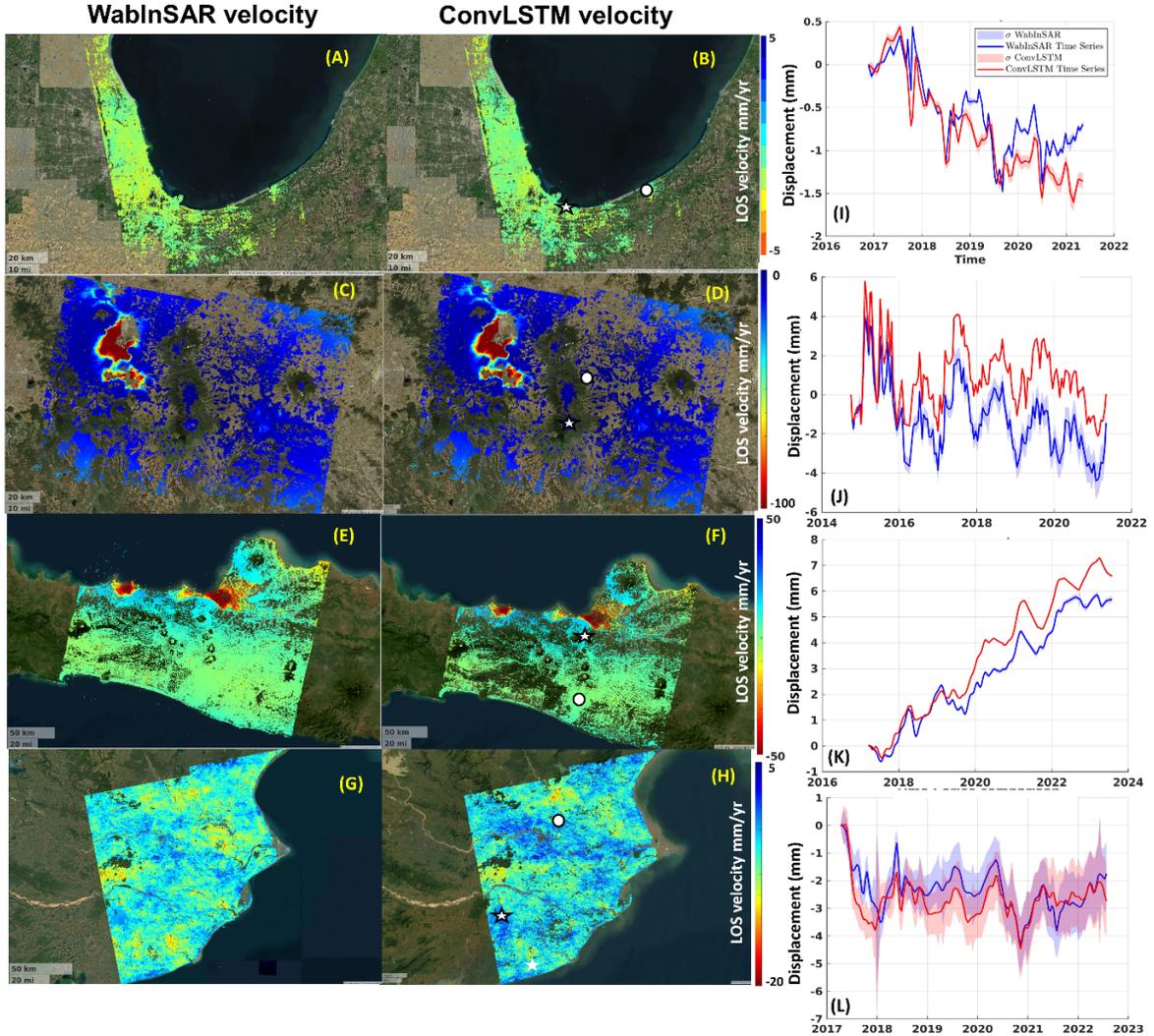

Fig. 4: Comparison of displacement maps generated from WabInSAR (first column) and CIPS model (second column). Panels (A), (C), (E) and (G) show 1D LOS velocity maps generated from original WabInSAR processing chain for Chicago, Mexico, Java Island and Indian east coast, respectively. Panels (B), (D), (F) and (H) show corresponding velocity maps for Chicago, Mexico, Java Island and Indian east coast, respectively, with modified WabInSAR chain replacing original pixel selection with CIPS model-based selection. White pentagon and white oval mark the test pixel and reference pixel locations respectively, used for comparison. Exact location of the test and reference points are shown in Supplementary Table VIII. Colorbars are common for each predicted dataset (e.g. A and B). Panels (I), (J), (K) and (L) show time series comparison plots for test points shown in (B), (D), (F) and (H).

## V. CONCLUSION

We propose and demonstrate CIPS, a deep learning-based architecture to speed up and efficiently perform elite pixel selection in time series InSAR. The models trained separately for urban and coastal sites (also encompassing non-urban terrain) learn from the spatio-temporal behavior of the interferometric phase and efficiently select elite pixels, outperforming the conventional InSAR processors in terms of computational efficiency, generating pixel selection results in order of minutes. The utility of the architecture is demonstrated by embedding it into an existing InSAR processing (WabInSAR) chain, finding that the quality of the final displacement architecture is retained. Compared to the traditional methods for pixel selection, the proposed model does not require probabilistic or statistical modelling, and is easy to implement even by non-domain experts. Further, the architecture can be easily integrated with any InSAR processing chains, promising a quick selection of both PS and DS pixels.

## ACKNOWLEDGMENT

The authors are grateful to the ESA for providing Sentinel-1 IW images. AT and LO are supported by NSF and NASA. MS and NS are supported by DOE.

## OPEN RESEARCH

The code and an example dataset used for training and testing the method developed in this study is available at the following link:

https://data.lib.vt.edu/articles/dataset/Deep_learning_for_efficient_selection_of_measurement_pixels_in_multi-temporal_InSAR_processing/23478236

## AUTHOR CONTRIBUTIONS

Ashutosh Tiwari: Conceptualization, Methodology, Data curation, Implementation, Validation, Writing – original draft. Nitheshnirmal Sadhasivam: Data preparation, Writing – editing original draft. Leonard O. Ohenhen- Data preparation, Writing – editing original draft. Jonathan Lucy- Writing- review and editing. Manoochehr Shirzaei: Conceptualization, Methodology, Software, Supervision, Writing – review and editing.

**Leveraging power of deep learning for fast and efficient elite pixel selection in time series interferometry**

Ashutosh Tiwari*, Nitheshnirmal Sadhasivam, Leonard O. Ohenhen, Jonathan Lucy and Manoochehr Shirzaei

Department of Geosciences, Virginia Tech, Blacksburg, VA, USA

Virginia Tech National Security Institute, Blacksburg, VA, USA

**Contents of this file**

Tables 1 to 5.

Figures S1 to S2.

**Introduction**

We provide supporting information to the main text related to the datasets used for training the proposed deep learning architecture, its configuration and training parameters. We also add a few tables related to interpretation of the results obtained from the proposed architecture.

TABLE I A. Urban datasets used for training and testing.

| Type | Dataset | Path | Frame(s) | Orbit | No. of Images | Temporal coverage | Training/Prediction time (mins) |
|---|---|---|---|---|---|---|---|
| Training (Urban) | LA | 64 | 108 | ASC | 79 | 2015-2021 | 124 |
| | NY | 33 | 130 | ASC | 81 | 2017-2021 | 121 |
| | Phoenix | 20 | 103, 108 | ASC | 79 | 2014-2021 | 134 |
| | Houston | 34 | 90, 95 | ASC | 82 | 2016-2021 | 138 |
| Testing (Urban) | Chicago | 121 | 130,135 | ASC | 80 | 2016-2021 | 10.5 (prediction) |
| | Mexico City | 143 | 526, 531 | DES | 137 | 2014-2021 | 2.5 (prediction) |

TABLE I B. Coastal datasets used for training and testing.

| Type | Dataset | Path | Frame(s) | Orbit | No. of Images | Temporal coverage | Training/Prediction time (mins) |
|---|---|---|---|---|---|---|---|
| Training (Coastal) | Chesapeake Bay | 4 | 115 | ASC | 116 | 2016-2020 | 124 |
| | Miami, Port St. Lucie | 48 | 79, 84 | ASC | 101 | 2015-2020 | 121 |
| | Boston, Yarmouth | 62 | 133 | ASC | 116 | 2016-2020 | 134 |
| | Southport, Wilmington | 77 | 107 | ASC | 112 | 2017-2020 | 138 |
| | Rhode Island, Brookhaven | 135 | 131 | ASC | 115 | 2016-2020 | 136 |
| | Savannah, Richmond Hill | 150 | 102 | ASC | 124 | 2015-2020 | 132 |
| Testing (Coastal) | Indian East Coast | 65 | 83 | ASC | 156 | 2017-2022 | 3 (prediction) |
| | Java Island coast | 76 | 613, 618 | DES | 167 | 2017-2023 | 3 (prediction) |

TABLE II. CIPS model configuration. Total trainable and non-trainable parameters are 93281 and 32 respectively. The terms $n_s$ and $n_t$ represent the number of image samples and time steps respectively.

| Layer (type) | Parameters | Output dimension |
|---|---|---|
| ($convlstm_1$) + *layernorm* + *relu* | No. of filters: 16 | ($n_s$, $n_t$,100,100,16) |
| ($convlstm_2$) + *batchnorm*+ *relu* | No. of filters: 16 | ($n_s$, $n_t$,100,100,16) |
| (*conv*+BN)2+relu | No. of filters: 16 | ($n_s$, 100,100,16) |
| *dropout* | Ratio: (0.2,0.25, 0.3) | ($n_s$, 100,100,16) |
| *FC* | | ($n_s$,100,100,1) |

TABLE III. Hyperparameters used for training the proposed models.

| Hyper-parameter | Values |
|---|---|
| Loss function | $f_1$ *loss* |
| Performance metric | $f_1$ *score, accuracy, precision, recall* |
| Training epochs | 20 to 200 epochs per dataset |
| Early stopping criteria | 5 epochs |
| Optimizer | Adam |
| Learning rate | (0.01, 0.001) |
| Decay rate | ($10^{-5}$, $10^{-4}$) |

TABLE IV. Comparison of pixel densities generated from WabInSAR and CLSTM-ISS model.

| S.No. | Site | WabInSAR pixel density | CLSTM-ISS pixel density | Prediction time by CLSTM-ISS (seconds) |
|---|---|---|---|---|
| 1 | Chicago | 6.32 | 7.31 | 624 |
| 2 | Mexico | 33.40 | 32.23 | 134 |
| 3 | Indian East Coast | 46.82 | 49.88 | 185 |
| 4 | Java Island coast | 21.66 | 21.16 | 172 |

TABLE V. Confusion matrices for test sites.

| Chicago | Predicted | Actual | |
|---|---|---|---|
| | | Positive | Negative |
| | Positive | 13913433 (TP) | 208560 (FP) |
| | Negative | 58896 (FN) | 893961 (TN) |

| Mexico | Predicted | Actual | |
|---|---|---|---|
| | | Positive | Negative |
| | Positive | 1401500 (TP) | 36240 (FP) |
| | Negative | 61475 (FN) | 659709 (TN) |

| Indian coast | Predicted | Actual | |
|---|---|---|---|
| | | Positive | Negative |
| | Positive | 1268420 (TP) | 166269 (FP) |
| | Negative | 83776 (FN) | 1179527 (TN) |

| Java coast | Predicted | Actual | |
|---|---|---|---|
| | | Positive | Negative |
| | Positive | 4585244 (TP) | 153773 (FP) |
| | Negative | 183806 (FN) | 1126922 (TN) |

TABLE VI. Prediction scores for the proposed CIPS model.

| S.No | Metric | Chicago | Mexico | Indian East Coast | Java Coast |
|---|---|---|---|---|---|
| 1 | Accuracy | 98.22 | 95.47 | 90.73 | 94.41 |
| 2 | Precision | 81.08 | 94.79 | 87.64 | 87.99 |
| 3 | Recall | 93.81 | 91.47 | 93.36 | 85.97 |
| 4 | F1-score | 86.98 | 93.10 | 90.41 | 86.97 |

TABLE VII: Testing urban model over coastal site and vice versa

| Study area | Type | WabInSAR | ConvLSTM (urban model) | ConvLSTM (Coastal model) |
|---|---|---|---|---|
| Java Island | Coastal | 1310728 | 539751 | 1280695 |
| Chicago | Urban | 952857 | 1102521 | 1721252 |

TABLE VIII: Test and reference points for the test datasets

| S.No | Site | Reference point (Longitude, Latitude) in decimal degrees | Random test point (Longitude, Latitude) |
|---|---|---|---|
| 1 | Mexico | (-98.53, 19.25) | (-98.62, 19.01) |
| 2 | Chicago | (-87.67, 41.80) | (-87.38 41.62) |
| 3 | Java Island | (110.37, -7.76) | (110.44, -7.15) |
| 4 | Indian East Coast | (86.27, 20.96) | (85.83, 20.27) |

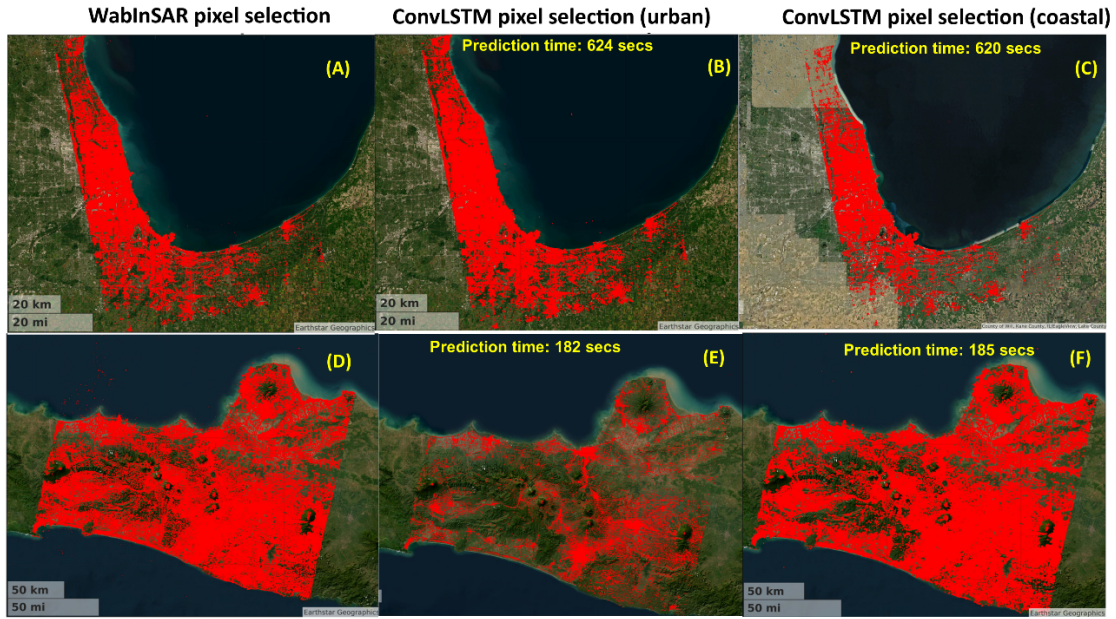

Fig. S1. Pixel selection results using coastal model on urban dataset and vice versa. (A), (B) and (C) show elite selected by WabInSAR, convLSTM run using urban model, and convLSTM run using coastal model

respectively, for Chicago (urban area). (D), (E) and (F) show pixels selected by WabInSAR, convLSTM run using urban model, and convLSTM run using coastal model respectively, for Java Island (coastal site).

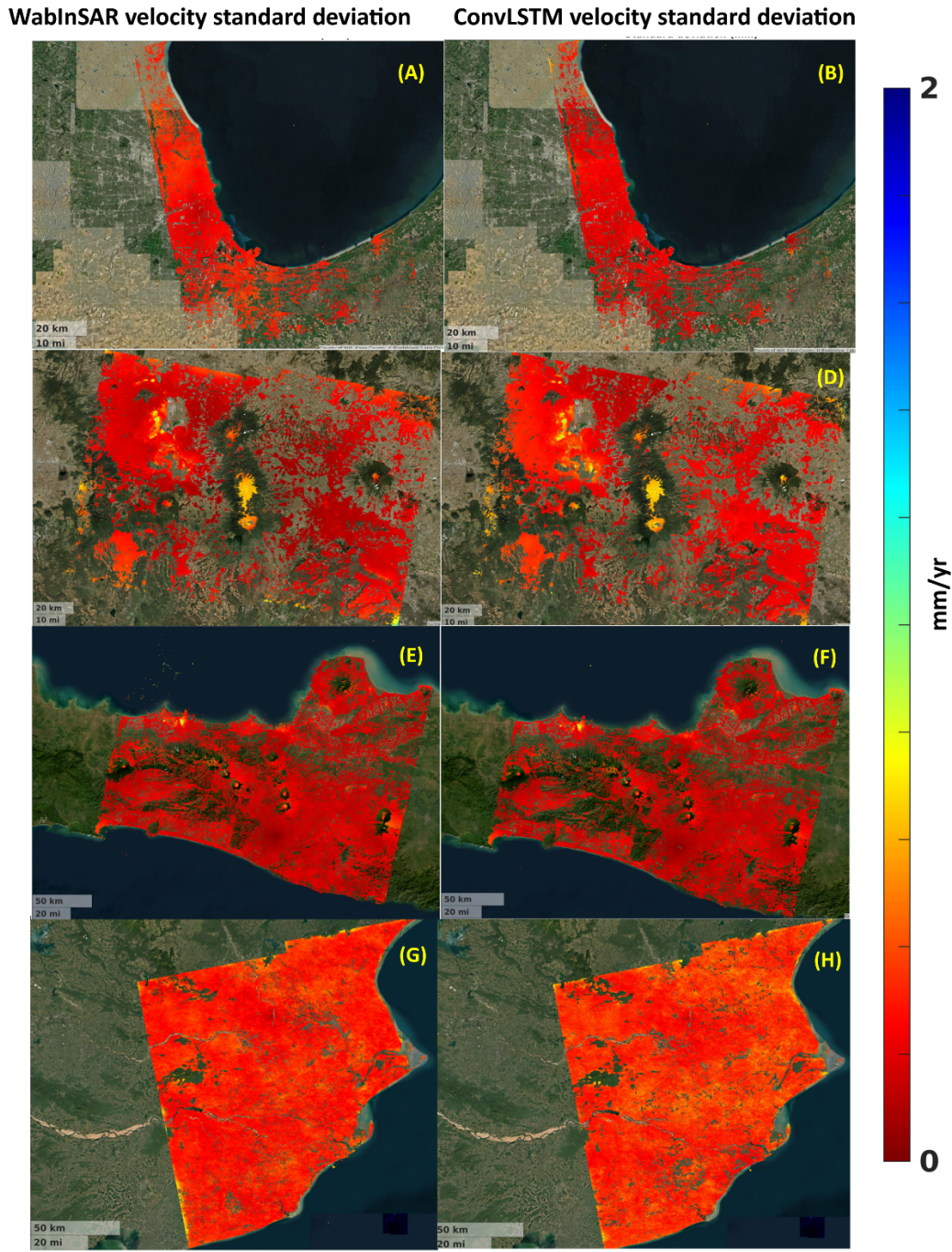

Fig. S2: Comparison of velocity standard deviations for mean velocity maps derived from conventional TS-InSAR software (WabInSAR) and the proposed Convolutional LSTM model. The Left and right columns show the results for the predicted urban and coastal sites with WabInSAR and ConvLSTM, respectively. Panels (A), (C), (E) and (G) show WabInSAR-based

velocity standard deviation maps for Chicago, Mexico, Java Island and Indian east coast, respectively. Panels (B), (D), (F) and (H) show corresponding CIPS model-based selection for Chicago, Mexico, Java Island and Indian east coast, respectively.

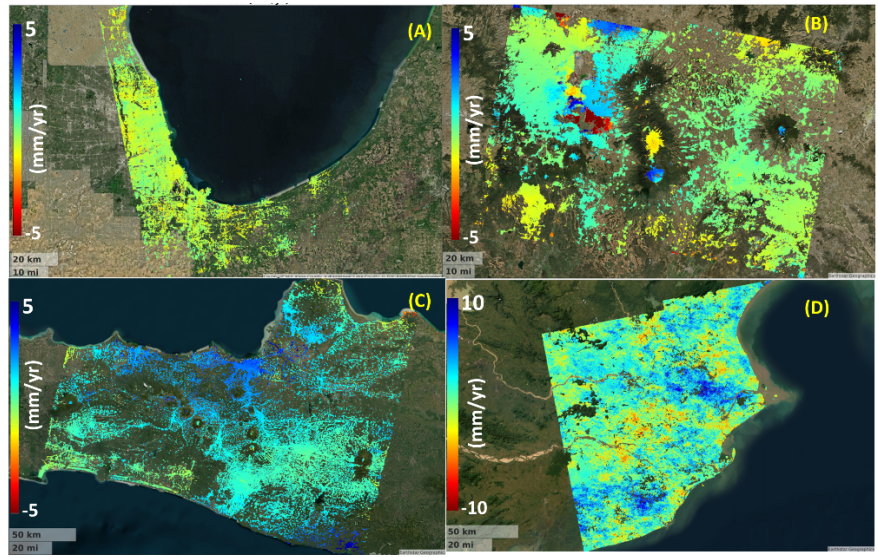

Fig. S3. Difference of mean line of sight velocities for the test sites. (A), (B), (C), and (D) show velocity difference maps for Chicago, Mexico, Java Island and Indian East Coast datasets respectively.